\documentclass[twocolumn,aps, amsmath,amssymb]{revtex4}

\usepackage{graphicx}
\usepackage{bm}

\usepackage[varg]{txfonts}

\newcommand{\be}{\begin{equation}}
\newcommand{\ee}{\end{equation}}

\newcommand{\laL}{\lambda_{\rm L}}
\newcommand{\veps}{\varepsilon}
\newcommand{\eps}{\epsilon}

\newcommand{\adu}[1]{a^\dag_{#1 \uparrow}}
\newcommand{\add}[1]{a^\dag_{#1 \downarrow}}

\begin{document}

\title{Period Doubling in  Small Multiply Connected Superconductors}

\author{Victor Vakaryuk}
\email[]{vakaryuk@uiuc.edu}

\affiliation{Department of Physics, University of Illinois, Urbana, Illinois 61801, USA}


\begin{abstract}
It is shown that for superconductors with circumference $2\pi R$ approaching BCS coherence length $\xi_0$ minimal period of the response of all thermodynamic quantities to external magnetic field is set by $hc/e$ i.e.~twice the corresponding value for the bulk case.
 This is explained by the dependence of internal energy of Cooper pairs on their center of mass motion which leads, in particular, to a transition offset between different current-carrying states. Explicit calculation  of the transition offset is done for the case of $s$-wave superconducting cylinder with $R \gg \xi_0$ and turns out to be exponentially small. A possible enhancement of the effect for nodal superconductors is suggested. 
Similar conclusions should also apply to the response of charged or neutral superfluids to rotation.
\end{abstract}


\maketitle

It is well known that response of  superconductors to electromagnetic field  and inertial perturbations such as rotation involves such characteristics of  Cooper pairs as their mass $2m_{\rm e}$  and charge  $2e$ with $m_{\rm e}$ and $e$ being essentially, apart from tiny relativistic corrections, 
electron's bare mass and charge.
 Although for all known superconductors the interparticle spacing is actually smaller (usually much smaller) than the size of the pairs, notions  of pair's mass and charge have  good  heuristic value in constructing phenomenological expressions for various thermodynamic and transport quantities 
 on the basis of their counterparts in Bose superfluids by a simple replacement $m \rightarrow 2m$ and $e \rightarrow 2e$.
 
The implications of this prescription on a quantum of superfluid circulation have been experimentally confirmed on a variety of physical systems where internal degrees of freedom of Cooper pairs are either absent or irrelevant. As an example of the electromagnetic response one can quote  classical flux quantization experiments in both conventional \cite{DeaverFairbankPRL1961, DollNabauerPRL1961} and $d$-wave superconductors \cite{Gough1987short}; inertial properties of Cooper pairs were tested, in particular, in observation of quantized circulation in $^3$He-B \cite{He3D, He3H} as well as in high precision measurements 
of magnetic field induced in a superconductor by rotation \cite{TatePRL1989}. While there are no known charged Bose superfluids so that the unit of flux quantization established in \cite{DeaverFairbankPRL1961, DollNabauerPRL1961, Gough1987short}, namely $hc/2e$, does not have its Bose counterpart, experiments with rotating superconductors \cite{TatePRL1989} do have their analog--the Hess-Fairbank effect--in rotating $^4$He and seem to well establish the circulation unit of $h/m_{\rm He}$ for $^4$He  \cite{HessPRL1967, LeggettBook} and $h/2m_{\rm e}$ for niobium in Ref.~\onlinecite{TatePRL1989} thus confirming to the prescription indicated above.

On the theoretical side 
flux quantization 
in superconducting samples of annular geometry has been attributed to the condensation of \emph{pairs} of particles into 
states with different momenta of the center of mass of the pairs
\cite{ByersPRL1961, BrenigPRL1961}. In thermodynamic equilibrium,  as the external field changes, discrete sequential transitions between
these states 
 lead to 
flux-periodic dependencies of \emph{all} quantities characterizing the annulus, e.g.~to a periodic dependence of induced magnetization.
Even without detailed knowledge of the ground state  and relying only on the gauge invariance principle one can conclude that the fundamental period of such dependencies will be $hc/e$ \cite{ByersPRL1961, BlochPRB1970}. However if the  ground state possesses pair correlations of the type mentioned above with all or most pairs being in the same  center of mass state then the response of the system will contain substantial $hc/2e$ harmonic rendering the minimal flux period of the dependencies to the same value. 

It has been noted a long time ago that there is no fundamental reason 
behind minimal flux periodicity being $hc/2e$
\cite{ByersPRL1961, BlochPRB1970}. 
Indeed calculation of the response of  an $s$-wave ring under the assumption of independence of relative motion of electrons on the flux \cite{Zhu1994} as well as calculations of Little-Parks effect in $s$-wave rings \cite{Bogachek:1975, Tzu-Chieh2007} and magnetization of mesoscopic $d$-wave loops \cite{Loder2008, Juricic2008}
have observed absence of $hc/2e$ periods in the corresponding quantities. The simple argument presented here shows that this situation, namely 
doubling of the minimal period
in the response of all thermodynamic quantities, is quite generic and happens irrespective of the particular form of pairing, interactions, temperature effects etc.~\emph{when the circumference  of the 
superconductor becomes small}. 
 For an $s$-wave superconductor where pair wave function (WF) decays exponentially, the effect becomes noticeable when the circumference  
approaches
BCS coherence length $\xi_0$. However for a nodal superconductor, such as $d$-wave,
 there are directions in the real space where decay of the pair WF is \emph{algebraic} which suggests that the effect  may be noticeable at even larger than $\xi_0$ values of the circumference. Similar analysis 
 goes through for rotating superconductors where the response is periodic in the rotation velocity with the minimal `unbroken' period equal to $\hbar/2 m_{\rm e}$.

Basic geometry considered in this work is that of a hollow cylinder with the wall thickness $d$ being smaller than the London penetration depth $\lambda_{\rm L}$ and the radius of the cylinder $R \gg d$; same conclusions can be reached for an arbitrary geometry requiring only axial symmetry and $d \ll \lambda_{\rm L}$ condition. Single particle states are specified by a set of three quantum numbers $(m, \bm{n})$ with $\hbar m$ being a projection of angular momentum along the symmetry axis and $\bm{n} \equiv (n_1,n_2)$ describing two other degrees of freedom responsible for the motion along and perpendicular to the symmetry axis. 

In thermodynamic equilibrium periodicity in the response of $N$ paired fermions is attributed  to  the transitions between  ground states of the following type
\be
	\Psi^{(m_0)}
	\!
	=
	\!
	\left[
	\frac 2 N
	\sum_{\bm{n} m} 
	\chi^{(m_0)}_m 
	(\bm{n}, \overline{\bm{n}})
	\,\,
	\adu{\bm{n}, m } \,
	\add{\overline{\bm{n}}, -m + {m_0}}
	\right]
	^{N/2}
	\hspace{-15pt}
	|0\rangle
\label{GSWF}
\ee
which describe condensation of $N/2$ spin-singlet pairs with the  pair's angular momentum along the symmetry axis equal to $\hbar m_0$.  Pairing with respect to the other two quantum numbers $\bm{n}$ is chosen in a standard way and connects state $\bm{n}$ with its time reversal $\overline{\bm{n}}$. Spin singlet symmetry of two-particle state  requires that variational parameter $\chi^{(m_0)}$ 
satisfies  $\chi^{(m_0)}_{-m +m_0} (\bm{n}, \overline{\bm{n}}) =  \chi^{(m_0)}_{m }  (\overline{\bm{n}}, \bm{n})$. 
The same condition is obeyed by the pair wave function in the momentum representation $F^{(m_0)}$ defined as
\be
	F^{(m_0)}_m 
	(\overline{\bm{n}}, \bm{n})
	=
	\langle
	a_{\overline{\bm{n}}, -m + m_0 \downarrow} 
	a_{\bm{n}, m \uparrow}
	\rangle
\label{pairWFmspace}
\ee
where the quantum mechanical average is taken between the states (\ref{GSWF}) with $N/2 - 1$ and $N/2$ pairs.  
	\begin{figure}
	\includegraphics[scale=0.55]{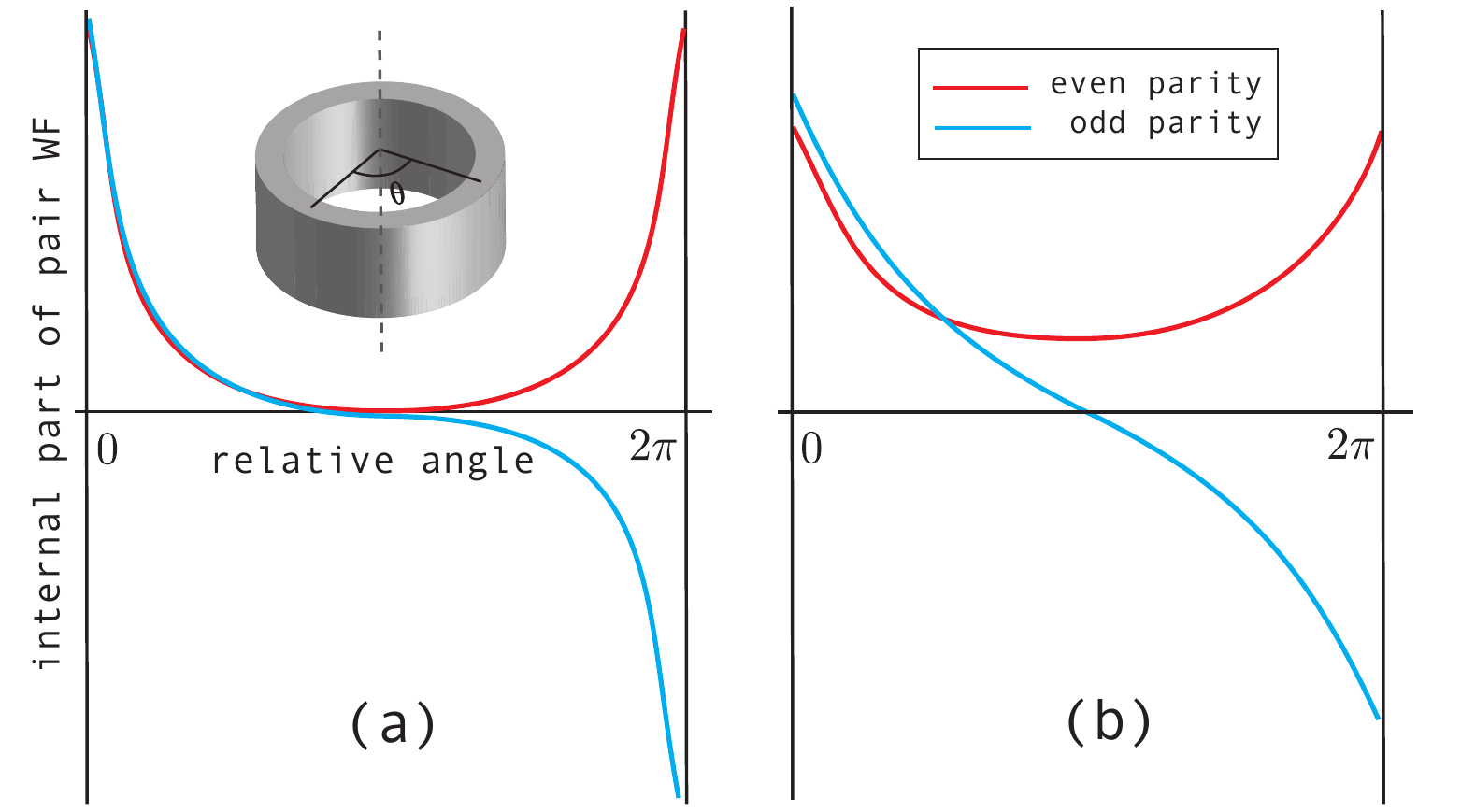}
	\caption{\label{fig:pairWF} \emph{Schematic} behavior of the internal part of  pair WF as a function of the relative angular coordinate for 	even (red) and odd (blue) values of the pair's angular momentum. High frequency oscillations are not shown. (a)  The circumference of the 	cylinder is much larger than the decay length of the pair WF. (b) The circumference of the ring is comparable with the decay length.
	\vspace{-10pt}
	}
\end{figure}
Structure of the pair WF is more apparent in the coordinate representation where it factorizes into two parts which describe 
internal motion and the motion of the center of mass of the pair:
\be
	F
	(\bm{x}_1 \theta_1; \bm{x}_2  \theta_2) 
	= 
	e^{i\frac{m_0}{2}(\theta_1 + \theta_2)} \,
	f^{(m_0)}
	(\bm{x}_1, \bm{x}_2, \theta_1 - \theta_2)
\label{pairWFrspace}
\ee
Here $\theta_i$ is a polar coordinate of $i$-th particle and $\bm{x}_i$ represents two other coordinates conjugate to the quantum number $\bm{n}$. Pair WF must be single valued in both $\theta_1$ and $\theta_2$ and symmetric under particles permutation $\bm{x}_1  \theta_1\rightleftarrows \bm{x}_2  \theta_2$. The former condition implies that internal part of the pair WF, $f^{(m_0)}(\bm{x}_1, \bm{x}_2, \theta)$, must be $2 \pi$ periodic (antiperiodic) for even (odd) parity values of $m_0$ as a function of the relative coordinate $\theta \equiv \theta_1 - \theta_2$.

Similar to (\ref{pairWFrspace}), center of mass decomposition can also be written for the coordinate representation of the variational parameter $\chi^{(m_0)}$ which suggests that 
energy expectation value taken on state (\ref{GSWF})
 can be factorized as well. As will be shown below this is indeed the case. Assuming a uniform magnetic flux $\phi$ through the annulus, the general form of a spin-independend single particle spectrum in axially symmetric geometry with thin walls is 
\be
	\veps_{\phi} (m,\bm{n})
	 =
	 \frac
	  	{\hbar^2}
		{2m_{\rm e}R^2}
	\left[
	  \zeta( \bm{n}) 
	  + 
	 (m + \phi / \phi_0)^2
	 \right]
\label{dispersion1}
\ee
where $\phi_0$ is the \emph{single} particle flux quantum $\phi_0 \equiv hc/|e|$ and $\zeta(\bm{n})$ is some dimensionless function. For such single-particle dispersion, independent of a particular form of interparticle interactions,  the expectation value of the ground state  energy evaluated on the state (\ref{GSWF})  can be represented as
\be
	E^{(m_0)}(\phi) 
	=
	\frac
		{N \hbar^2}
		{2 m_{\rm e} R^2}
	\left[
	\veps^{(m_0)}
	+
	(m_0/2 + \phi / \phi_0)^2
	\right]
\label{Ephi1}
\ee
In the expression above dimensionless parameter $\veps^{(m_0)}$ represents flux-independent  contribution  coming from the internal motion of the particles in the pair. Gauge invariance demands same parity values of $\veps^{(m_0)}$ to be the same which allows one to introduce two in general independent constants,
$\veps^{(0)} \equiv  \veps^{(2 m)}$ and
$\veps^{(1)} \equiv \veps^{(2 m+1)}$, 
describing even and odd parity states.

Another contribution to the system's energy  comes from the motion of the center of mass of the pair 
 and is represented by the second term 
 on the r.h.s.~of eqn.~(\ref{Ephi1}). It is usually assumed that the two contributions 
  are independent, namely, that the energy associated with the internal structure of the Cooper pair is independent of the motion of the center of mass  of the pair. It is the violation of this assumption which leads to the breaking of $hc/2e$ periodicity in small samples. 

To see why for small rings the internal energy of the Cooper pair depends on its center of mass state it is instructive to turn to the real space representation of the pair WF (\ref{pairWFrspace}) whose internal part is \emph{schematically} shown on fig.~\ref{fig:pairWF} for $\bm{x}_1 = \bm{x}_2$.  For the circumferences much larger than the characteristic decay length of the pair WF fig.~\ref{fig:pairWF}a applies and the internal energy being independent on the overall phase of WF is practically the same for odd and even parity states. Then the circumference becomes comparable with the characteristic decay length, the continuity condition for the pair WF requires it to adjust as schematically shown on fig.~\ref{fig:pairWF}b
	\footnote{
			There are two implicit assumptions required for the validity of this statement. In the first place, the decay length should 				not shrink 	as fast as the circumference itself. Secondly, the ring should be 			uniform enough to avoid  a possibility of placing kinks in the pair WF in the regions where its magnitude is suppressed.
			}. 
The internal energies evaluated on such states are \emph{different}.

In the simplest case of $s$-wave superconductor the decay of the pair WF in real space is isotropic and exponential with the decay length 
equal to
the BCS coherence length $\xi_0$. The internal energy difference for this case will be evaluated below under assumption $R \gg \xi_0$ and will turn out to be exponentially small. However for a superconductor with a nodal structure in the gap there are directions in the real space where pair WF decays \emph{algebraically} which may lead to an enhancement of the effect.

Coming back to eqn.~(\ref{Ephi1}) and introducing superfluid velocity $v_s$ and superfluid density $\rho_s$ for the \emph{pairs} of particles which occupy volume $\Omega$ by usual relations $v_s \equiv \frac{\hbar}{2 m_{\rm e} R} (m_0 + \frac{\phi}{\phi_0/ 2})$, $\rho_s \equiv m_{\rm e} N / \Omega$ 
it becomes 
\be
	\Omega^{-1}
	E^{(m_0)}(\phi) 
	=
	\frac
		{\hbar^2}
		{2 m_{\rm e}^2 R^2}
	\,
	\rho_s
	\veps^{(0,1)}
	+
	\frac
		{1}
		{2}
	\,
	\rho_s v_s^2
\label{Ephi2}
\ee
Being written in this form the expression for energy can, in fact, be generalized to include effects of non-zero temperature or other pair breaking perturbations by introducing corresponding changes in the superfluid density $\rho_s$ as well as other terms describing e.g.~behavior of the excited component of the system. Such generalizations will not change \emph{qualitatively} conclusions reached below and the  following discussion will be limited to the simplest case specified by eqn.~(\ref{Ephi2}).

\begin{figure}[t]
\includegraphics[scale=0.55]{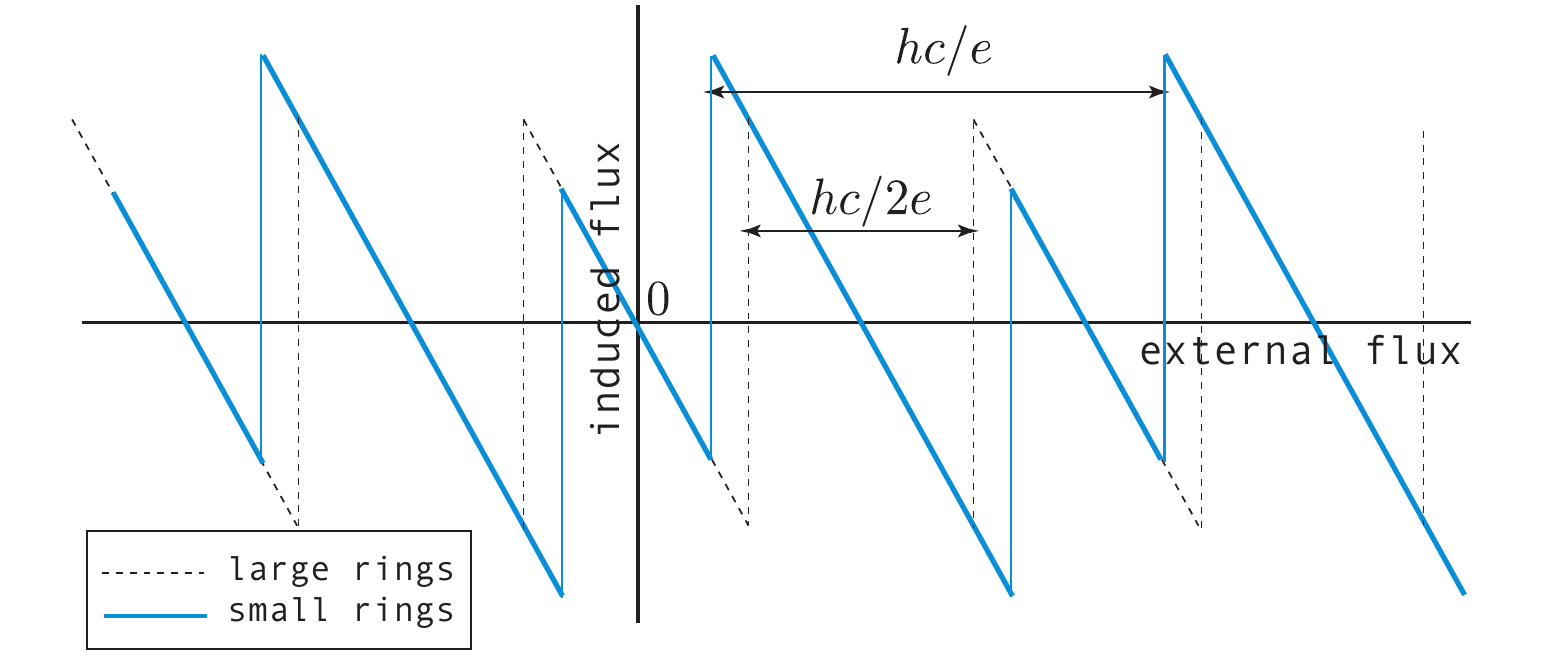}
\caption{\label{fig:extFlux} Dependence of the induced flux on the external flux for rings of small and large sizes.
}
\vspace{-10pt}
\end{figure}

The discussion so far has led to the expressions (\ref{Ephi1}) and (\ref{Ephi2}) for the ground state energy as a function of pair's angular momentum $\hbar m_0$ and total flux $\phi$.  To find thermodynamically stable value of $m_0$ which is realized at a fixed \emph{external} flux $\phi_e$ one needs to construct corresponding thermodynamic potential which is in this case Gibbs potential $G$.
 Although general expression for $G$ can be easily written down, 
 for the purpose of this discussion it is enough to notice that, in  thermodynamic equilibrium, the transition between $m_0=0$ and $m_0=1$ states occurs at the following value of the external flux:
\be
	\phi_e^{(0 \rightarrow 1)}
	=
	\frac 1 2 
	\,
	\frac{hc}
		{2e}
	\left[
	1 + 
	\alpha
	\left(
	{\veps}^{(1)}
	-
	{\veps}^{(0)}
	\right)
	\right]
\label{phi0}
\ee
where ${\veps}^{(0,1)}$ is the dimensionless internal energy of the Cooper pair for the states $m_0=0,1$ defined by eqn.~(\ref{Ephi1}). Appearance of parameter $\alpha \equiv 1+  4 R d / \laL^2$ in (\ref{phi0}) has purely geometrical reasons  related to fact that, for the geometry considered here, field and condensate energies scale differently with $R$. 
As one can see from the result above, transition between different 
center of mass
 states is offset relative to the bulk case by an amount proportional to the internal energy difference.

Complete dependence of the induced flux obtained by minimization of the Gibbs potential corresponding to the energy (\ref{Ephi1}) is illustrated on fig.~{\ref{fig:extFlux}}. The effect of the non-vanishing internal energy difference is to shift the transitions between different 
current
 states by an amount proportional to $\pm ({\veps}^{(1)}-{\veps}^{(0)})$ thus breaking $hc/2e$ periodicity up to $hc/e$
	\footnote
			{
			It should be pointed out that this mechanism of breaking $hc/2e$ periodicity is different from that proposed in Ref.~\cite				{Loder2008} where it is the behavior of the excited component of the superfluid that breaks the periodicity. The position 				of the the transitions between  different 
			center of mass 
			states is unaffected in that model.
			}. 
In an alternative interpretation one may refer to this effect as a change in Cooper pair's charge. A very similar reasoning shows that the same change occurs for the Cooper pair mass in the rotating superconductors thus leaving gyromagnetic ratios like $2m_{\rm e} c /e$ unchanged.

The preceding qualitative conclusions were reached without particular reference to a microscopic form of interparticle interactions.
However for a quantitative analysis, which is most easily done in the momentum representation, a choice of specific microscopic model is needed. Using  single-particle dispersion relation (\ref{dispersion1}) and a contact interparticle interaction potential $V$ energy  expectation value on a many-body ground state (\ref{GSWF}) is
\begin{eqnarray}
	\hspace{-10pt}
	E^{(m_0)}(\phi)
	\!\!
	 &=&
	 \!\!\!
	  \sum
	2 \veps_{\phi}(m, \bm{n})  
	|v_{m-m_0/2}^{(m_0)}(\bm{n}, \overline{\bm{n}})|^2
	\nonumber
	\\
	&-&
	\!\!\!
	 \sum
	V \,
	{\cal F}_{m-m_0/2}^{\,(m_0)}(\bm{n}, \overline{\bm{n}})  
	{\cal F}_{m'-m_0/2}^{\,(m_0)*}(\bm{n}' \! , \!  \overline{\bm{n}}')
\label{varEnergy}
\end{eqnarray}
where quantities $v$ and $\cal F$ are defined through earlier introduced $\chi$, eqn.~(\ref{GSWF}), and $F$, eqn.~(\ref{pairWFmspace}), as 
$|v_{m-m_0/2}^{(m_0)}(\bm{n}, \overline{\bm{n}})|^2 \equiv  \chi^{(m_0)}_m  (\bm{n}, \overline{\bm{n}}) F^{(m_0)*}_m  (\bm{n}, \overline{\bm{n}})$
and
${\cal F}_{m-m_0/2}^{\,(m_0)}(\bm{n}, \overline{\bm{n}}) \equiv F^{(m_0)}_m  (\bm{n}, \overline{\bm{n}})$. 
Defined in such way $v$ and $\cal F$ are \emph{symmetric} with respect to $m$, e.g.~$v_m  (\bm{n}, \overline{\bm{n}}) = v_{-m}  ( \overline{\bm{n}}, \bm{n})$ etc. In what follows, to avoid complicated notation,   $\bm{n}$ dependence of all quantities will not be indicated explicitly unless mentioned otherwise. 

Noticing that normalization condition for state (\ref{GSWF}) requires $2 \sum |v_{m-m_0/2}^{(m_0)}|^2 \!\!=\! \!N$,  the flux dependence in (\ref{varEnergy}) can be separated out thus establishing a connection with the expression for energy (\ref{Ephi1}) 
used in the previous analysis. Identifying flux dependent part with the 
the center of mass contribution
and everything else with the internal energy contribution one thus obtains for the latter
\begin{eqnarray}
	N \,
	{\veps}^{(m_0)}
	&=&
	2
	\sum 
	(
	{\zeta}
	+
	(m -m_0/2)^2
	)
	|v_{m-m_0/2}^{(m_0)}|^2
	\nonumber
	\\
	&-&
	\!
	{V}
	\sum
	{\cal F}_{m-m_0/2}^{\,(m_0)}
	{\cal F}_{m'-m_0/2}^{\,(m_0)*}
\label{internalmspace}
\end{eqnarray}
and the interaction constant is now written in units of $\hbar^2/2m_{\rm e} R^2$. It is also apparent that ${\veps}^{(m_0)}$ defined by the equation above is the same for the same parity states, as was indicated above.

The actual forms of $v^{(m_0)}$ and ${\cal F}^{(m_0)}$ can be found by energy minimization in the same way as it is done in the standard BCS treatment which leads to
$|v^{(m_0)}_m|^2 = |v_m(\Delta^{(m_0)} \!\!, \mu^{(m_0)})|^2$, 
${\cal F}_{m}^{(m_0)} = {\cal F}_{m}(\Delta^{(m_0)} \!\!, \mu^{(m_0)})$ 
where, for a given pair state $m_0$,  the gap parameter, $\Delta^{(m_0)}$,  and the chemical potential, $\mu^{(m_0)}$, should be determined self-consistently from gap and normalization equations  and are, in general, different for odd and even parity states. For fixed $\Delta^{(m_0)}$ and $\mu^{(m_0)}$ functions $v^{(m_0)}$ and ${\cal F}^{(m_0)}$ are given by standard BCS expressions.

In the momentum representation the reason for  a non-vanishing  value of the internal energy difference between  odd 
  and even
    parity states and, as a result, doubling of the flux periodicity, is less transparent than in the coordinate representation but from a mathematical point of view it can be traced down to two main factors.  
In the first place, the actual values of $m$ arguments which $v$ and $\cal F$ are being summed over in (\ref{internalmspace}) are shifted by $1/2$ for odd and even parity cases; secondly, there is a difference between odd and even values of the gaps and chemical potentials. It can be shown that the latter difference being itself caused by the former in the gap and normalization equations can be ignored in the leading approximation. The actual calculation of the energy difference can be found in the Appendix. Performing summation over $m$ with the help of Poisson summation formula and integrating afterwards with respect to two other quantum numbers $\bm n$ it is, in fact,  possible to get an analytic expression for ${\veps}^{(1)} - {\veps}^{(0)}$ in the limit $R \gg \xi_0$. With the relative accuracy $\xi_0/R$ one has
\be
	{\veps}^{(1)}
	 -
	{\veps}^{ (0)}
	 =
	 \frac
		{3 }
		{\pi^{3/2}}
	 	{
		\left[
		\frac
			{R}{ \xi_0}
		\right]
		^{1/2}
		}
	\! e^{-2 {R}/\xi_0}
	\cos (2 \pi k_{\rm F} R)
\label{result}
\ee
where $k_{\rm F}$ is the Fermi wave vector and $\xi_0$ is defined through the gap parameter $\Delta$ as $\xi_0 \equiv \hbar v_{\rm F}/ \pi \Delta$. It should be emphasized that  the accuracy of eqn.~(\ref{result}) does not allow one to distinguish between $\Delta^{(1)}$ and $\Delta^{(0)}$ (or $\mu^{(1)}$ and $\mu^{(0)}$) since the difference $\Delta^{(1)} - \Delta^{(0)}$ is itself exponentially small. There might be however a substantial difference between either of $\Delta^{(0,1)}$ and the value of the gap for the corresponding \emph{bulk} material.  

One should observe that the sign of eqn.~(\ref{result})  is a rapidly oscillating function of the radius which is due to the oscillations of the pair WF in real space. Similar oscillatory  effects  have  been found theoretically in Ref.~\cite{BlattPRL1963} for the dependence of the gap on the thickness of a thin superconducting film. 
Strong dependence of the sign of the transition offset on the ring's size
would make it more difficult to observe it on an ensemble of rings, such as the one used in \cite{Koshnick:Moler2007} for Little-Parks measurement, because of possible variations in the rings' sizes. 

Another limitation for experimental observation is the magnitude of the effect. According to eqn.~(\ref{result}) one may expect transition offset to be of the order of 0.1\% for $R \approx 3 \xi_0$;  larger corrections for the same value of $R/\xi_0$ will most likely be achieved for nodal superconductors. Although reducing the radius will make the effect more pronounced, a little more care should be taken in extrapolating eqn.~(\ref{result}) to the ring sizes equal or lesser of the coherence length. This is due to the fact that when the diameter of the ring becomes equal to the coherence length the superfluid velocity needed to screen a quantum of flux reaches its critical value thus leading to the destruction of superconductivity around $\phi_e = \phi_0 /2$ \cite{Liu2001}.

It is interesting to notice that, apart from the small sample size, there might be other  ways to introduce correlations between the center of mass and internal energies of the Cooper pair.
One of them could be a strong spin-orbit interaction where the small size limitation would not apply.

To conclude, it was suggested that in small superconductors, due to the dependence of internal energy of Cooper pairs on the center of mass state, the minimal flux periodicity is $hc/e$, twice the usually attributed value. 
The doubling of the periodicity is due to the offset of the transition between different current states.
The magnitude of such offset was calculated for $s$-wave pairing. It was also suggested that other things being equal, the effect will be more pronounced for nodal superconductors.

The author wishes to thank Tony Leggett for guidance and David Ferguson and Alexey  Bezryadin for useful discussions. Support from NSF grant DMR03--50842 is gratefully acknowledged.


\appendix*
\section{}

In this appendix expression (\ref{result}) for the internal energy difference between odd and even parity states of a large, $R \gg \xi_0$, $s$-wave superconducting cylinder or ring will be obtained. Unless explicitly otherwise stated all quantities with dimensions of energy are written in units of $\hbar^2 /2m_{\rm e} R^2$.

As it has been already remarked  the minimization of the expectation value of energy given by eqn.~(\ref{varEnergy})  leads to the the following form of  $v_{m}^{(m_0)}$ and ${\cal F}_{m}^{(m_0)}$:
\begin{eqnarray}
	|v^{(m_0)}_m|^2 
	&=&
	 |v_m(\Delta^{(m_0)} \!\!, \mu^{(m_0)})|^2
	 \\
	{\cal F}_{m}^{(m_0)} 
	&=& 
	{\cal F}_m (\Delta^{(m_0)} \!\!,  \mu^{(m_0)})
\end{eqnarray}
where, for given values of $\Delta$ and $\mu$, $v_m$ and ${\cal F}_m$ are specified by the standard BCS expressions:
\begin{eqnarray}
	|v_m|^2 
	&=&
	\frac 1 2 
	\left[
	1- \frac{\epsilon_m}{\sqrt{\eps_m^2 + \Delta^2}}
	\right] 
\label{v}
	\\
	{\cal F}_m
	&=& 
	 \frac{\Delta}{2\sqrt{\eps_m^2 + \Delta^2}}
\label{F}
\end{eqnarray}
with $\eps_m \equiv \zeta +m^2 -\mu$  being a single-particle energy counted from  the chemical potential which corresponds to a given value of $m_0$.

Now, using eqn.~(\ref{internalmspace}), the internal energies for even and odd parity states are given by
\be
	N
	\veps^{(0)}_0
	=
	2 \sum 
		 (\zeta +m^2)
		 |v_{m,0}|^2 
	 -
	V \sum
		{\cal F}_{m,0} {\cal F}_{m',0}^*
\ee
for even parity states, and
\begin{eqnarray}
	N
	\veps^{(1)}_1
	&=&
	2 \sum 
		 (\zeta +(m -1/2)^2)
		 |v_{m-1/2,1}|^2
	\nonumber
	\\
	 &-&
	V \! 
	\sum
		{\cal F}_{m-1/2,1} {\cal F}_{m'-1/2,1}^*
\end{eqnarray}
for odd parity states. The subscript 0 or 1 in all quantities indicates dependence on  $\Delta^{(0)}$, $\mu^{(0)}$ or $\Delta^{(1)}$, $\mu^{(1)}$ respectively. The internal energy difference can then be written as
\be
	\veps^{(1)}_1
	-
	\veps^{(0)}_0
	=
	\veps^{(1)}_1 
	-
	\veps^{(0)}_1 +\delta \veps^{(0)}
\label{e1e0}
\ee 
where $\delta \veps^{(0)} \equiv \veps^{(0)}_1- \veps^{(0)}_0$ is the change of the ground state energy ($m_0 = 0$ state) as  $\Delta$ and $\mu$ are varied from their $m_0=0$ to $m_0=1$ values. Since $\veps^{(0)}_0$ is the equilibrium i.e.~\emph{minimal} value of $\veps^{(0)}$,  the last term on the r.h.s.~of eqn.~(\ref{e1e0}) is of the second order in $\Delta^{(1)} - \Delta^{(0)}$  and, as it will be seen below, can be ignored relative to the first two terms in the limit $R \gg \xi_0$.
 Thus to evaluate the  internal energy difference in the leading order the energies themselves can be taken at the same values of $\Delta$ and $\mu$. At this accuracy any difference between $\Delta^{(0)}$ and $\Delta^{(1)}$ etc.~is neglected; however, there might be a substantial difference between these values and the corresponding bulk parameters.

Using the gap equation $\Delta = V \sum {\cal F}_{m -1/2}$, in the leading approximation eqn.~(\ref{e1e0}) becomes
\begin{eqnarray}
	N
	(
	\veps^{(1)} 
	 \!\!\!
	&-&
	\veps^{(0)}
	)
	=
	2
	 \sum
	(\zeta + (m - 1/2)^2) 	v_{m-1/2}^2
	\\
	&-&
	\sum
	(\zeta + m^2) 	v_{m}^2
	- 
	2 \Delta
	\sum
	(
	{\cal F}_{m - 1/2} - {\cal F}_m
	)
	\nonumber
\end{eqnarray}
where $v_m$ and ${\cal F}_m$ given by eqn.'s (\ref{v}) and (\ref{F}). The sums above run over all integers $m$ as well as over two other quantum numbers $\bm{n}$ which represent other-than-azimuthal part of the dispersion $\zeta(\bm{n})$. With the help of the Poisson summation formula the sum over $m$ is  converted to an integral;  summation over $\bm{n}$ is replaced by integration using the standard rule $\sum_{\bm{n}} \rightarrow \int \! d\zeta \, g_2(\zeta)$ where $g_2(\zeta)$ is the (dimensionless) density of states for $\zeta(\bm{n})$ and can be considered to be a constant.
Keeping only leading exponential term in the Poisson summation series one obtains:
\be
	N
	(
	\veps^{(1)} 
	 \!\!
	-
	\veps^{(0)}
	)
	=
	I
	+
	I_1
\label{ediff}
\ee
where the following definitions are made
\be
	I
	\equiv
	4 \mu
	\int
	 \! d\zeta \, g_2
	\int 
	\! dx \,
	\frac
		{\eps(x, \zeta) \, e^{2\pi i x}}
		{\sqrt{\eps^2(x, \zeta) + \Delta^2}}
\ee
and
\be
	I_1
	\equiv
	\!
	4
	\!\!
	\int
	 \!\! d\zeta \, g_2
	 \!\!
	\int 
	\!\! dx \!
	\left[
	{\sqrt{\eps^2(x, \zeta) + \Delta^2}}
	-
	\eps(x, \zeta)
	\right]
	\, e^{2\pi i x}
\ee
with $\eps(x, \zeta) \equiv \zeta +x^2 - \mu$. It is intuitively plausible and can, in fact, be shown that $I_1 / I \sim \xi_0 / R \ll 1$ so that the problem reduces to the calculation of the following integral:
\be
	I
	\equiv
	4 \mu
	\int_{-\mu}^\infty
	\!\!
	g_2
	 \, d\zeta \, 
	\int_{-\infty}^\infty
	\! dx \,
		\frac
		{({\zeta} + x^2) \, e^{2\pi i x}}
		{\sqrt{({\zeta} +x^2)^2+\Delta^2}} 
\label{2piiz}
\ee
where and from now on $\zeta$ will be counted \emph{relative to the chemical potential}. The rest of the discussion is devoted to the calculation of the above integral under assumption  $R \gg \xi_0$.

The integration over $x$ in eqn.~(\ref{2piiz}) is transformed by closing the integration contour in the upper half of the complex plane where the denominator $\sqrt{({\zeta} +x^2)^2+\Delta^2}$ has two branch points. Choosing the branch cuts to point outwards away from the origin  the integral over the real axis reduces to four integrals along both sides of the two branch cuts. The two integrals along one branch cut are complex conjugate of those along the other branch cut. Introducing new integration variable $t$ along the first quarter branch cut by $z = z_0 (t+1)$, where $z_0$ is the first quarter branch point, the oscillating exponent in (\ref{2piiz}) acquires  a decaying part:
\be
	I
	=
	16 \mu \,  \Re
	\int_{-\mu}^\infty
	\!\!\!
	g_2
	\, d \zeta 
	\!
	\int_{0}^\infty
	\! \! \!\!  z_0 dt \,
		\frac
		{({\zeta} + z_0^2(t+1)^2) \, e^{2\pi i z_0(t+1)}}
		{\sqrt{({\zeta} +z_0^2(t+1)^2)^2+\Delta^2}} 
\label{2piiz0}
\ee
where $\Re$ denotes the real part of the corresponding expression and the decaying exponent is given by the imaginary part of $z_0$:
\begin{eqnarray}
	\Re \, z_0
	\equiv
	x_0 
	&=&
		\frac{1}
			{\sqrt{2}} 
		\left(
		- {\zeta} 
		+\sqrt{{\zeta}^2 +\Delta^2}
		\right)^{1/2}
	\label{x0}
	\\
	\Im \, z_0
	\equiv
	y_0 
	&=&
		\frac{1}
			{\sqrt{2}} 
		\left(
		 {\zeta} 
		+\sqrt{{\zeta}^2 +\Delta^2}
		\right)^{1/2}
	\label{y0}
\end{eqnarray}

The essential for the integration over $t$ region in (\ref{2piiz0}) is determined by the decaying exponent $e^{-2\pi y_0 t}$ and extends from $t_{\rm min} = 0$ to $t_{\rm max} \sim (2\pi y_0)^{-1}$. As a function of ${\zeta}$, $y_0$ is monotonically increasing  in the region $(-\mu, \infty)$ reaching its minimum at ${\zeta} = -\mu$ with the value 
$y_0(-\mu) \approx  {\Delta}/{2 \sqrt{\mu}}$; at the same time $x_0$ is monotonically decreasing with the maximum value
$x_0(-\mu) \approx \sqrt{\mu}$
	\footnote
			{Usual energy scale separation $\mu \gg \Delta$ is assumed.
			}.
Restoring dimensions i.e.~supplying $\hbar^2/2m_{\rm e}R^2$ denominators to $\Delta$ and $\mu$ gives 
\be
	y_0(-\mu)
	 \approx  
	{R}/{\pi \xi_0},
	\quad
	x_0(-\mu)
	\approx
	k_{\rm F} R
\ee
where $\xi_0$ is the BCS coherence length $\xi_0 \equiv \hbar v_{\rm F} / \pi \Delta$ and $v_{\rm F}^2 \equiv 2\mu/m_{\rm e}$
\footnote
	{
	Numerical value of the parameters can be estimated using 
	$
	\sqrt {\frac {2 m_{\rm e} R^2} {\hbar^2}} \frac {\Delta} {2 \sqrt{\mu}} 
	\approx
	 R
	 (\mu\text{m}) \,
	  \frac
	  	{\Delta (\text{K})}
		{\sqrt{\varepsilon_{\rm F} (\text{eV})}}
	$.
	}.

The above analysis suggests that, under condition $R \gg \xi_0$, the integrand in (\ref{2piiz0}) can be significantly simplified because in the important for the integration region $t \ll 1$ holds.  After some algebraic manipulations followed by a Wick rotation of  $t$ variable to make the algebraic part of the integrand real for $\zeta < 0$, the expression for the integral takes the following form
\be
	I
	=
	8\mu \,  \Re
	\int_{-\mu}^\infty 
	\!\!\!
	g_2 d\zeta
	\!
	 \int_{0}^\infty
	\!\!\! i
	\,  dt \,
		\frac
			{(- 2 t {\zeta} +  \Delta) \, e^{2\pi i z_0(it+1)}}
			{\sqrt
				{
				t \, ( \Delta - {\zeta} t )
			}	} 
\label{2piiz0it}
\ee
with further corrections 
being suppressed by a factor of order of $\xi_0/R$.

The integration over $\zeta$ is considered separately for intervals of positive and negative values of $\zeta$. After rescaling variable $t$ by $\Delta / |\zeta|$ in each interval the non-trivial dependence of the integrand on $\Delta$ and $\zeta$ is gathered in the exponent. For the integral over  $\zeta > 0$ region the exponent has a sharp maximum and the integral can be evaluated by the saddle point method leading to its magnitude being of order of $\exp{(-2\pi \sqrt{\Delta})}$. As it will be seen below the integral over negative values of $\zeta$ is of order of $\exp{(-2\pi \Delta/ \sqrt{\mu}})$. Since the ratio of the the two exponents is $\sqrt{\Delta/\mu} \ll 1$ the integration over $\zeta$ in (\ref{2piiz0it}) can be limited to the interval $\zeta <0$ leading to
\be
	I
	=
	8 \mu \,  
	\Delta
	\int_{-\mu}^0
	\frac
		{
		g_2 d \zeta
		}
		{
		\sqrt{|{\zeta}|}
		}
	\int_{0}^\infty
	\!  dt \,
		\frac
			{( 2  t  +  1) \,  \sin Y}
			{\sqrt
				{
				t \, ( 1 + t )
			}	} 
		\, e^{-  X}
\label{mX}
\ee
where 
\begin{eqnarray}
	X({\zeta}, t) 
	&\equiv&
		2 \pi y_0({\zeta}) + x_0({\zeta}) \frac{2 \pi \Delta}{|{\zeta}|} t
	\\
	Y({\zeta}, t)
	&\equiv&
			-2 \pi x_0({\zeta}) + y_0({\zeta})  \frac{2 \pi \Delta}{|{\zeta}|} t
\end{eqnarray}
with $x_0$ and $y_0$ given by eqn.'s (\ref{x0}) and (\ref{y0}). 

Expressions for $X({\zeta}, t)$ and $Y({\zeta}, t)$ can be significantly simplified in the limit $|\zeta| \gg \Delta$. In particular for $\zeta = -\mu$
\begin{eqnarray}
	X(-\mu, t) 
	&\approx& 
	4 \tilde{R} \,
	(1/2 +t)
	\label{Xmu}
	\\
	Y(-\mu, t)
	&\approx&
	-2 \pi \sqrt{\mu}
	+
	\frac{\Delta}{\mu} \,
	2 \tilde{R}
	\, t
	\label{Ymu}
\end{eqnarray}
where $\tilde{R}$ is defined as
\be
	\tilde{R}
	\equiv
	\pi \Delta / \sqrt{\mu}
\ee
and in dimensional units is  equal to $R/ \xi_0$, the ratio of the radius of the annulus to the coherence length. As it has been already mentioned it is assumed that $\tilde{R} \gg 1$.

The function in the exponent in eqn.~(\ref{mX}), $X({\zeta}, t)$, is a positive monotonically growing  function of ${\zeta}$ for ${\zeta} < 0$ which reaches its minimum at the lower integration limit $\zeta = -\mu$ and diverges at the upper limit ${\zeta} = 0$. To take the advantage of the simplified forms of $X$ and $Y$ at $\zeta = - \mu$ one can perform repeated integration by parts in the integral over $\zeta$ obtaining in that way following asymptotic expansion:
\begin{widetext}
\be
	I
	=
	8g_2 \mu \Delta
	\int_0^\infty
	\!\! dt \,
	\frac
		{2t + 1}
		{\sqrt{t(t+1)}}
	\left. 					
	\left\{
	Z(\zeta, t)
	+
	\left[
		\frac{Z(\zeta, t)}{ X'(\zeta, t)}
	\right]'
	+
	\left[
		\frac{1}{ X'(\zeta, t)}
		\left[
			\frac{Z(\zeta, t)}{ X'(\zeta, t)}
		\right]'
	\right]'
	+
	\ldots
	\right\}
	\right|_{-\mu}
	\frac
		{e^{- X(-\mu, t)}}
		{ X'(-\mu, t)} 
\label{expansion1}
\ee
\end{widetext}
where the prime sign denotes differentiation with respect to $\zeta$ and function $Z(\zeta, t)$ is defined as the $\zeta$ dependent non-exponential part of the integrand in (\ref{mX}):
\be
	Z(\zeta, t)
	\equiv
	\frac
		{1}
		{\sqrt{|\zeta|}}
	\,
	\sin Y(\zeta, t)
\label{Zmu}
\ee
Despite complicated at first sight $t$ dependence of the integrand in (\ref{expansion1}) the integration over $t$ can be carried out noticing that, because of the decaying exponent, the interval relevant to the integration is limited by $\tilde{R}^{-1} \ll 1$.  Dropping $t$ dependence relative to constants of order 1 and integrating the expansion  term by term gives with $\tilde{R}^{-1}$ relative accuracy
\be
	I
	=
	4 g_2 \mu^2 \Delta
	\,
	\pi^{1/2}
	\,
	S( \mu)
	\, 
	\tilde{R}^{-3/2}
	e^{-2 \tilde{R}} \,
\ee
where $S(\mu)$ is defined as 
\be
	S(\mu)
	\equiv
	\left. 					
	\left\{
	Z(\zeta, 0)
	+
	\left[
		\frac{Z(\zeta, 0)}{ X'(\zeta, 0)}
	\right]'
	+
	\ldots
	\right\}
	\right|_{-\mu}
\ee
The sum above cannot be limited to a finite number of terms because the $n$-th term in the sum is of order of $\mu^{-1/2} (\mu/ \Delta)^{n-1}$. However by rearranging summation one can notice that $S(\mu)$ satisfies following differential equation
\be
	S(\mu)
	=
	Z(-\mu, 0)
	-
	\frac
		{d}
		{d \mu}
	\,
	\frac
		{S(\mu)}
		{X'(-\mu, 0)}
\ee
or, given explicit form of $X(-\mu,0)$ and $Z(-\mu, 0)$, eqn.'s (\ref{Xmu}) and (\ref{Zmu}), the differential equation for $S(\mu)$ becomes
\be
	S(\mu)
	=
	-
	\frac{1}
		{\sqrt{\mu}}
	\sin{2\pi\sqrt{\mu}}
	-
	\frac{1}
		{\pi \Delta}
	\frac{d}
		{d \mu}
	\left[
	\mu^{3/2}
	S(\mu)
	\right]
\ee
It can be checked by a direct substitution that up to the terms of order of $\Delta/\mu$ this equation has the following solution:
\be
	S(\mu)
	=
	\frac{\Delta}
		{\mu^{3/2}}
	\,
	\cos{2\pi \sqrt{\mu}}
\ee
so that the expression for the integral $I$ reduces to
\be
	I
	=
	\frac
		{4 g_2}
		{\pi^{3/2}}
	\,\, \mu^{3/2}
	 \tilde{R}^{1/2}
	\, e^{-2 \tilde{R}}
	\,
	\cos 2 \pi \sqrt{\mu}
\ee
where $g_2$ is the density of states at the chemical potential for other-than-azimuthal part of the dispersion $\zeta(\bm{n})$. Using single-particle dispersion law for a thin walled cylinder and reverting to dimensional units one can see that $g_2 \mu^{3/2} = \frac 3 4 N$ so that
\be
	I
	=
	\frac
		{3N }
		{\pi^{3/2}}
	\,\,
	 (R/\xi_0)^{1/2}
	\, e^{-2 R/\xi_0}
	\,
	\cos 2 \pi  k_{\rm F} R
\ee
Comparing this equation with eqn.~(\ref{ediff}) gives the result (\ref{result}) for the internal energy difference quoted in the main text.


\end{document}